\documentclass[useAMS,usenatbib]{mn2e}

\usepackage{times}
\usepackage{graphicx}
\usepackage{xspace}
\usepackage{epsfig}
\usepackage{rotating}
\usepackage{dcolumn}

\title[FU\,Tau: accretion and activity]{The enigmatic young brown dwarf binary FU\,Tau: \\ accretion and activity}
\author[Stelzer et al.]{B. Stelzer$^{1}$\thanks{E-mail:
stelzer@astropa.inaf.it} and A. Scholz$^{2}$ and C. Argiroffi$^{1,3}$ and G. Micela$^{1}$\\
$^{1}$INAF - Osservatorio Astronomico di Palermo, 
  Piazza del Parlamento 1,
  I-90134 Palermo, Italy \\ 
$^{2}$   School of Cosmic Physics, 
  Dublin Institute for Advanced Studies,  
  31 Fitzwilliam Place, Dublin 2, Ireland \\
$^{3}$   Dipartimento di Scienze Fisiche ed Astronomiche, Sezione di Astronomia, 
  Universit\`a di Palermo, Piazza del Parlamento 1, I-90134 Palermo, Italy}
\begin{document}

\date{Accepted 2010 June 10.  Received 2010 June 8; in original form 2010 April 17}

\pagerange{\pageref{firstpage}--\pageref{lastpage}} \pubyear{2010}

\maketitle

\label{firstpage}

\begin{abstract}
FU\,Tau belongs to a rare class of young, wide brown dwarf binaries. 
We have resolved the system in a {\em Chandra} X-ray observation
and detected only the primary, FU\,Tau\,A.
Hard X-ray emission, presumably from a corona, is present but, unexpectedly, we   
detect also a strong and unusually soft 
component from FU\,Tau\,A. Its X-ray properties, so far
unique among brown dwarfs, are very similar to those of the T\,Tauri star TW\,Hya. 
The analogy with TW\,Hya suggests that the dominating
soft X-ray component can be explained by emission from accretion shocks. 
However, the typical free-fall velocities of a brown dwarf are too low for an interpretation
of the observed X-ray temperature as post-shock region. 
On the other hand, velocities in excess of the free-fall speed are derived from archival
optical spectroscopy, and 
independent pieces of evidence for strong accretion in FU\,Tau\,A are found in optical photometry.
The high X-ray luminosity of
FU\,Tau\,A coincides with a high bolometric luminosity confirming an unexplained trend among young brown dwarfs. 
In fact, FU\,Tau\,A is overluminous with respect to evolutionary models 
while FU\,Tau\,B is on the $1$\,Myr isochrone suggesting non-contemporaneous formation of the two
components in the binary. 
The extreme youth of FU\,Tau\,A could be responsible for its peculiar X-ray properties, 
in terms of atypical magnetic activity or accretion. Alternatively, rotation and magnetic
field effects may reduce the efficiency of convection which in turn affects 
the effective temperature and radius of FU\,Tau\,A shifting its position in the HR diagram. 
Although there is no direct prove of this latter
scenario so far we present arguments for its plausibility.

\end{abstract}


\section{Introduction}\label{sect:intro}

The young binary brown dwarf (BD) FU\,Tau has aroused interest in the star formation community 
for two particular aspects both having important implications for BD formation theories \citep[see][]{Luhman09.1}:
(1) It is one of only a handful of known binary BDs with wide separation ($800$\,AU). 
(2) It is located in the B\,215 dark cloud in Taurus that hosts only one other T\,Tauri star
(TTS). It is the only 
young ($<5$\,Myr) BD known in such a remote location and, therefore, it has 
formed most likely in isolation and not in a stellar cluster or aggregate. 
From low-resolution optical spectra 
\cite{Luhman09.1} measured spectral types of M7.25 and M9.25 for the two components FU\,Tau\,A and
FU\,Tau\,B. They derive masses of $0.05\,M_\odot$ and $0.015\,M_\odot$, respectively, comparing the
position of the two BDs in the HR diagram to pre-MS evolutionary models by \cite{Baraffe98.1} and \cite{Chabrier00.2}. 
They find that FU\,Tau\,A is significantly overluminous relative to the youngest ($1$\,Myr) isochrone of the
models. 
The extinction was measured from infrared (IR) 
photometry to be $\sim 2$\,mag for the primary and $< 1$\,mag for the secondary. 
The spectral energy distributions of both binary components indicate the presence of circumstellar disks
by excess emission over the photospheric model in the {\em Spitzer} mid-IR bands. 
In addition, there is a blue excess for FU\,Tau\,A that might indicate ongoing accretion. 
Both FU\,Tau\,A and~B have strong H$\alpha$ emission in low-resolution spectra 
(equivalent widths of $93$ and $70$\,\AA, respectively), again a likely accretion signature. 

Widely separated BDs are important calibrators of sub-stellar magnetic
activity because they allow to examine the X-ray characteristics of two presumably coeval targets with different 
(sub-)stellar properties ($T_{\rm eff}$, $L_{\rm bol}$, etc.) in a critical parameter range.
Magnetic activity (as probed by H$\alpha$ and X-ray luminosity) seems to decay strongly at late-M
spectral types \citep{Mohanty05.1, Stelzer06.1}. 
{\em Chandra} is the only X-ray instrument capable of resolving FU\,Tau and similar BD binaries. In an earlier study
we have resolved the wide BD binary in the Chamaeleon star-forming region, 2MASS\,J11011926-7732383\,AB
\citep{Stelzer07.3}. 
In a continuation of our search for coronal activity in sub-stellar twins 
we have obtained a {\em Chandra} X-ray observation of FU\,Tau. 
Here we combine the analysis of the X-ray data with archival optical spectroscopy and photometry. 
The primary turns out to have unexpected X-ray properties. This observation shows how incomplete our
knowledge of the high-energy emission of BDs is.

\section{Data analysis and results}\label{sect:data_analysis}

\subsection{X-ray observation}\label{subsect:data_xrays}

On 22 Oct 2009 FU\,Tau was observed for $55$\,ksec with {\em Chandra}/ACIS-S (Obs-ID\,10984). 
The data analysis was carried out with the CIAO 
package\footnote{CIAO is made available by the {\em Chandra} X-ray Center and can be downloaded 
from http://cxc.harvard.edu/ciao/download/}, version 4.1,   
we started with the level\,1 events file provided by the {\em Chandra} X-ray Center. 
The steps performed to convert these data to an event level\,2 file are 
described e.g. in \cite{Stelzer07.3}. 

\begin{figure}
\begin{center}
\includegraphics[width=8cm]{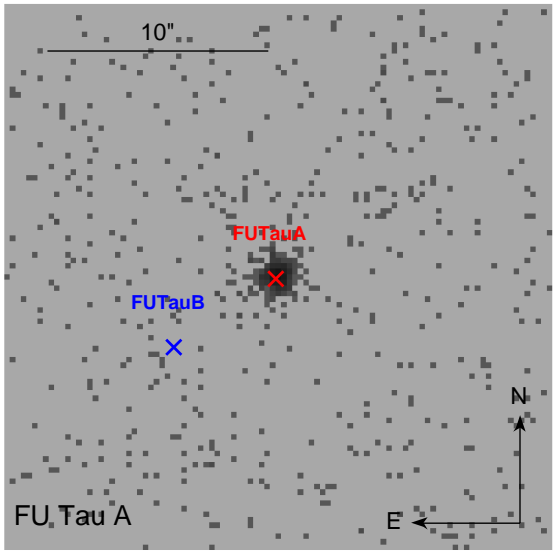}
\caption{{\em Chandra}/ACIS image of the FU\,Tau binary. 2\,MASS positions are marked with
crosses. The secondary is not detected.}
\label{fig:acis_image}
\end{center}
\end{figure}

The X-ray image is almost devoid
of X-ray sources, not unexpectedly as high extinction through the B\,215 cloud obscures the background sky.
However, bright X-ray emission is found to coincide with the 2\,MASS position of FU\,Tau\,A. 
For detecting the X-ray source(s) associated with the two components of the BD binary,  
source detection was restricted to a $100 \times 100$ pixels wide image 
(1\,pixel $= 0.5^{\prime\prime}$) and a congruent, monochromatic exposure map for $1.5$\,keV
centered on the 2\,MASS position of FU\,Tau\,A. 
Source detection was carried out with the {\sc wavdetect} algorithm  
with wavelet scales between $1$ and $8$ in steps of $\sqrt{2}$. 
We tested a range of detection significance thresholds, and found that $\sigma_{th} = 10^{-5}$ 
avoided spurious detections and at the same time separated close emission components. 
The X-ray image in the region around FU\,Tau\,A is shown in Fig.~\ref{fig:acis_image}. 
There is only one X-ray source 
in the surroundings of the BD binary.  
It is associated with FU\,Tau\,A. The companion FU\,Tau\,B remains undetected. 

We calculated the source count rates in the following way: 
A circular source photon extraction region was defined as the area that contains 
$95$\,\% of the point spread function at the position of FU\,Tau\,A. The background was
extracted individually from a squared region centered on the source extraction area and 
several times larger than the latter one. 
A circular area centered on the position of the X-ray source was excluded from the background area. 
The S/N was computed from the counts summed in 
the source and background areas, respectively, after applying the appropriate 
area scaling factor to the background counts. In practice, the background is very low (a fraction
of a count in the source extraction area). 
Finally, the count rate of FU\,Tau\,A was obtained using the exposure time at
the source position extracted from the exposure map. 
We have estimated a $95$\,\% confidence upper limit for the count rate 
at the position of FU\,Tau\,B using the algorithm of \cite{Kraft91.1}. 

In Table~\ref{tab:xrayparams} we summarize some X-ray parameters of the binary.
The distance assumed for the estimate of luminosities is $140$\,pc.
For FU\,Tau\,A we report the value of $L_{\rm x}$ extracted from its spectrum (see below),
while for FU\,Tau\,B we used PIMMS\footnote{The Portable Interactive Multi-Mission Simulator (PIMMS) is accessible at http://cxc.harvard.edu/toolkit/pimms.jsp} to convert the count
rate upper limit to a flux limit. This conversion depends on the unknown spectral parameters.
The value given in Table~\ref{tab:xrayparams} refers to an assumed $N_{\rm H} = 2 \cdot 10^{21}\,{\rm cm^{-2}}$
(corresponding to the upper limit measured for the $A_{\rm V}$ of FU\,Tau\,B) and $kT = 1$\,keV
(as typical for a coronal plasma). Note that the influence of $N_{\rm H}$ on the count-to-flux conversion
is much stronger than that of the temperature. 
\begin{table}\begin{center}
\caption{X-ray parameters of the FU\,Tau binary.}
\label{tab:xrayparams}
\begin{tabular}{lrrrrr}\hline
\multicolumn{1}{c}{Object} & \multicolumn{1}{c}{Offax}       & \multicolumn{1}{c}{Counts}           & \multicolumn{1}{c}{Expo}  & $\log{L_{\rm x}}$ & $\log{(\frac{L_{\rm x}}{L_{\rm bol}})}$ \\
                           & \multicolumn{1}{c}{[$^\prime$]} & \multicolumn{1}{c}{in $0.3-8$\,keV}  & \multicolumn{1}{c}{[sec]} & [erg/s]           &                                       \\ \hline
FU\,Tau\,A & $  0.29 $ & $   603.8 \pm   24.6$ & $  54153$ & $ 29.7$ &  $-3.2$ \\
FU\,Tau\,B & $  0.39 $ & $  <  4.7$            & $  54053$ & $<27.2$ &  $< -3.8$ \\
\hline
\multicolumn{6}{l}{Bolometric luminosities from \protect\cite{Luhman09.1}.} \\
\end{tabular}
\end{center}\end{table}

For FU\,Tau\,A 
a lightcurve was extracted and searched for variability with a maximum likelihood method that divides
the sequence of photons in intervals of constant signal \citep[see ][]{Stelzer07.1} and, independently, 
with the Kolmogorov-Smirnov (KS) test. 
There is no significant variability in the X-ray count rate of FU\,Tau\,A (KS-test probability is $0.34$), 
such that the bright emission can not be attributed to a flare. 

The high signal has allowed us to fit the X-ray spectrum of FU\,Tau\,A. 
An individual response matrix and auxiliary response were extracted for
the position of FU\,Tau\,A using standard CIAO tools. 
As mentioned above, the background of ACIS is negligibly low.
We fitted the spectrum, rebinned to a minimum of $15$ counts per bin, 
in the XSPEC\,12.5.0 environment 
with a one- or two-temperature thermal model subject to absorption.  
The spectrum of FU\,Tau\,A can be described by an absorbed $2$-T thermal model 
(see Table~\ref{tab:xspec} for 
the best fit parameters). 
For standard extinction
laws \citep[e.g.][]{Predehl95.1} the observed X-ray absorption, $N_{\rm H}$, is
compatible with the extinction of $A_{\rm V} = 2$\,mag measured in the IR.  

In summary, with more than $600$\,counts the {\em Chandra} observation of FU\,Tau\,A 
represents the highest quality X-ray data obtained so far for a (young) BD. 
The most remarkable fact is that the emission is dominated by a cool plasma component of temperature
$kT = 0.24$\,keV. 
The soft plasma has four times more emission measure (EM) than the 
hotter component that has a typical coronal temperature 
($1.1$\,keV). 
The X-ray luminosities of the cool and hot components are $\log{L_{\rm x,1}}\,{\rm [erg/s]}=29.5$ and 
$\log{L_{\rm x,2}}\,{\rm [erg/s]}=29.2$, respectively, at $0.3-8.0$\,keV.

\begin{table*}\begin{center}
\caption{X-ray spectral parameters of absorbed $2$-T APEC model for FU\,Tau\,A compared to those of TW\,Hya.}
\label{tab:xspec}
\begin{tabular}{lcrrrrrrr}\hline
Object     & Instrument                & $\chi^2_{\rm red}$ (dof) & $\log{N_{\rm H}}$  & $kT_1$ & $kT_2$ & $\log{EM_1}$       & $\log{EM_2}$       & $\log{L_{\rm x}}^*$ \\
           &                           &                          & [${\rm cm^{-2}}$]  & [keV]  & [keV]  & [${\rm cm}^{-3}$]  & [${\rm cm}^{-3}$]  & [erg/s]             \\ \hline
FU\,Tau\,A           &  {\em Chandra}/ACIS       & 0.9 ( 27)              &  $21.8^{21.9}_{21.5}$ &   $0.24^{0.34}_{0.19}$ &   $1.12^{1.28}_{0.99}$ & $52.9^{53.5}_{52.1}$  & $52.3^{52.4}_{52.2}$  & $29.5$  \\
TW\,Hya $^\dagger$   &  {\em XMM-Newton}/EPIC-pn & 2.3 (216)              &  $20.8$               &   $0.23$               &   $1.22$               & $53.0$              & $52.2$                & $29.8$  \\
\hline
\multicolumn{9}{l}{$^*$ Intrinsic X-ray luminosity in the $0.5-2.0$\,keV passband; distance assumed for TW\,Hya is $51$\,pc (Mamajek 2005)} \\
\multicolumn{9}{l}{$^\dagger$ The best-fit model of TW\,Hya has peculiar abundances (not listed here); abundances have been set for all elements} \\
\multicolumn{9}{l}{to $0.2$ solar for FU\,Tau\,A because of the low photon statistics.} \\
\end{tabular}
\end{center}\end{table*}

\subsection{Optical spectroscopy}\label{subsect:data_opt}

FU\,Tau\,A was observed with Gemini-North and GMOS in long-slit mode as part of 
science program GN-2008A-Q-94, PI Thomas Dall. The data were taken in March 2008 and are publicly 
available from the Gemini Science Archive. Two spectra have been obtained, one with grism R\,600 
(exposure time $1200$\,s, resolution $R\sim 4000$, March 1) and one with grism B\,1200 
($1200$\,s, $R\sim 4000$, March 16). In both cases, a 0\farcs5 slit has been used. 
The quoted values for the resolution are taken from the Gemini website. 
To our knowledge 
this source has never been observed with a higher spectral resolution. For comparison, the 
spectra used by \cite{Luhman09.1} have $R<1000$ (at $6000$\,\AA). 

\begin{figure}
\begin{center}
\resizebox{9cm}{!}{\includegraphics[width=8cm,angle=270]{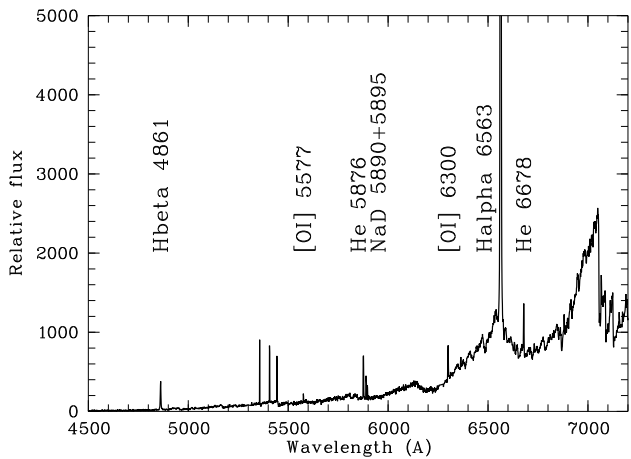}}
\caption{Optical spectrum of FU\,Tau\,A from Gemini/GMOS, 
dominated by H$\alpha$ emission and with several other prominent 
emission lines indicative of accretion and/or outflow. The three features between $5300$ and $5500$\,\AA~ 
are of telluric origin.}
\label{fig:speclow}
\end{center}
\end{figure}

We downloaded the science and calibration files from the Gemini archive and 
performed standard reduction using the Gemini/GMOS package in IRAF 
following 
the GMOS `cookbook'\footnote{http://www.gemini.edu/sciops/data/IRAFdoc/gmosinfospec.html}. 
This includes bias subtraction, flatfield correction, wavelength calibration, sky subtraction, 
and extraction. Since we are only interested in emission lines, we did not carry out a flux calibration. 
The final reduced R\,600 spectrum is shown in Fig.~\ref{fig:speclow} with the most prominent emission lines marked. 
The B\,1200 spectrum does not contain H$\alpha$ and has in general lower signal-to-noise ratio.

The H$\alpha$ line is broad with an equivalent width (EW) of $146$\,\AA, which is higher than the values 
published by \cite{Luhman09.1}. However, EWs are highly sensitive to the chosen value for the 
continuum and the integration range \citep{Stelzer07.2}.
For a rough assessment of the H$\alpha$ strength in FU\,Tau\,A and its comparison to other
objects we calculate the $L_{\rm H\alpha}/L_{\rm bol}$ ratio. We avoid using 
the stellar radius in our estimate because of its uncertainties in the case of 
FU\,Tau\,A (see discussion in Sect.~\ref{subsect:disc_hrd}). 
Estimating the continuum flux from the model and 
with $F_{\rm bol} = \sigma T^4_{\rm eff}$ for the bolometric flux, we find 
$L_{\rm H\alpha}/L_{\rm bol} = EW \cdot F_{\rm cont}/F_{\rm bol} \sim 2 \cdot 10^{-3}$.

We measure a $10$\,\% width \citep{Mohanty05.1} 
of $352\,{\rm km/s}$ for the H$\alpha$ line. 
The H$\alpha$ profile is not well-resolved, therefore this value has to be 
seen as upper limit. Still, both the EW and the $10$\,\% width are clearly above the typically adopted 
thresholds between non-accretors and accretors of $10$\,\AA~ and $200\,{\rm km/s}$. Moreover, 
FU Tau A features a variety of other emission lines which are often seen in accretors,  
e.g. H$\beta$, He\,5876, He\,6678 \citep{Jayawardhana06.3}. 
Thus, the optical spectroscopy clearly confirms FU\,Tau\,A as an accreting object. 
The [OI]\,6300 and NaD feature may also be related to outflow/accretion activity, 
but contamination by telluric emission cannot be excluded. 

From the $10$\,\% width and based on the correlation given by \cite{Natta04.2} we derived an 
accretion rate of $\dot{M}_{\rm H\alpha} = 3.5 \cdot 10^{-10}\,{\rm M_\odot/yr}$. Another estimate for the accretion 
rate can be obtained based on the line flux in He\,5876 and the correlation published by 
\cite{Herczeg08.1}. In this line, we measure EWs of $10$\,\AA~ and $11$\,\AA~ in the two spectra, i.e.
no significant variability is present. Scaling with the 
continuum flux, which is obtained from the AMES-Dusty model spectrum for $T_{\rm eff} = 2800$\,K and $\log{g} = 3.5$ 
\citep{Allard01.1}, this gives a line flux of $\log{F_{5876}}\,{\rm [erg/cm^2/s]} = 5.4$, 
which translates into an accretion rate of $\dot{M}_{\rm He\,I} = 7.5 \cdot 10^{-10}\,{\rm M_\odot/yr}$. 
Adopting Eq.~8 of \cite{Gullbring98.1},  
the observed $\dot{M}_{\rm He\,I}$ can be converted into the accretion luminosity, $L_{\rm acc, He\,I}$. 
With an assumed inner disk truncation radius $R_{\rm in} = 2\,R_*$ 
the measured $\dot{M}_{\rm He\,I}$ corresponds to $L_{\rm acc, He\,I} = 1.2 \cdot 10^{30}$\,erg/s. 
Our estimate assumes zero veiling in the continuum 
\citep[see][]{Mohanty05.1}; the actual accretion rate could therefore be slightly higher. 
Note, that $\dot{M}_{\rm He\,I}$ is higher than $\dot{M}_{\rm H\alpha}$ but compatible within the uncertainties given
by the empirical relations ($\sim 0.5$ logarithmic dex). The agreement with the accretion rates determined from
He\,$5876$ indicates that the large $10$\,\%
width of the H$\alpha$ line is intrinsic, i.e. the line is probably resolved.

\section{Discussion}\label{sect:discussion}

\subsection{X-ray emission from young BDs}\label{subsect:disc_bds}

A systematic survey of X-ray emission from BDs in the Taurus star-forming region
has been performed by \cite{Grosso07.1}. They used data from the {\em XMM-Newton Extended Survey
of the Taurus Molecular Clouds} \citep[XEST;][]{Guedel07.3} that included
$17$ young BDs. An X-ray detection fraction of $53$\,\% was obtained
for this sample. 
Only a small subset of the XEST BDs had enough photons collected ($> 100$\,counts) 
to enable a spectral analysis. For most of them the statistics was so poor that the spectral shape
could be represented adequately with a moderately absorbed ($N_{\rm H} < 10^{22}\,{\rm cm^{-2}}$) 
$1$-T thermal model of $kT \sim 0.5...1.5$\,keV, typical for the plasma in a magnetically heated corona.

For none of the BDs detected in XEST the X-ray emission is 
nearly as soft as for FU\,Tau\,A, despite they all have similar or 
lower absorbing column favoring the detection of a cool plasma component. 
Recall, that in FU\,Tau\,A the X-ray $EM$ and flux of the cooler spectral component exceeds that of the
hotter one (see Sect.~\ref{subsect:data_xrays}).
Dominating soft X-ray plasma has been detected so far only in older BDs, 
e.g. the evolved BD Gl\,569\,B has an emission measure weighted mean temperature of 
$kT = 0.6$\,keV at an age of $100$\,Myr \citep{Stelzer04.1} 
and the $500$\,Myr-old LP\,944-20 had $0.3$\,keV  \citep{Rutledge00.1}. Both of them were detected during a flare, 
i.e. in an event known to go along with plasma heating, and their quiescent emission -- if any -- is probably
even softer. The decrease of $L_{\rm x}/L_{\rm bol}$ and $kT$ of BDs with age has tentatively been explained by 
reduced coronal heating related to the increasingly cool effective temperatures 
\citep{Stelzer06.1}. 

In the $0.5-8$\,keV band, used by \cite{Grosso07.1} for XEST, 
the X-ray luminosity of FU\,Tau\,A ($4 \cdot 10^{29}\,{\rm erg/s}$) 
is at the high end of that of all BDs in Taurus. 
\begin{figure}
\begin{center}
\includegraphics[width=9cm]{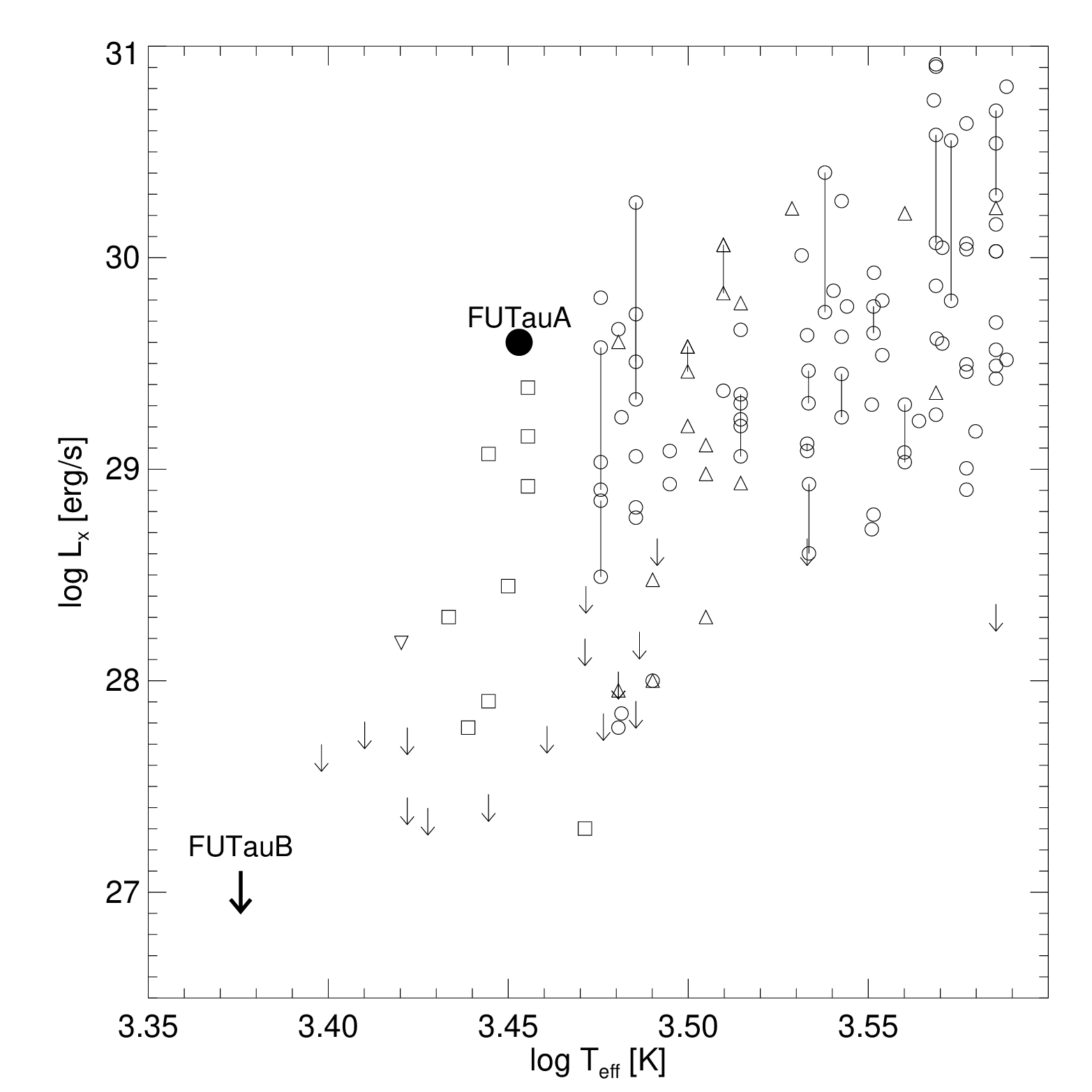}
\caption{X-ray luminosity versus effective temperature for young stars (circles -- \protect\cite{Guedel07.3},
upward pointing triangles -- \protect\cite{Scelsi07.1}) and 
BDs (squares -- \protect\cite{Grosso07.1}, downward pointing triangle -- \protect\cite{Mokler02.1}) 
in Taurus and the position of FU\,Tau. 
Non-detections in X-rays are indicated by downward pointing arrows.
The vertical lines connect the X-ray emission levels for stars observed and detected in two XEST pointings. 
The luminosities for the TTS given by \protect\cite{Guedel07.3} and shown in this figure 
refer to a slightly broader energy band ($0.3-10$\,keV)
but for a typical coronal X-ray spectrum ($kT \sim 1$\,keV) 
the difference in flux to the $0.5-8.0$\,keV band used in the BD study is negligible.}
\label{fig:lx_teff}
\end{center}
\end{figure}
%
In Fig.~\ref{fig:lx_teff} we show the X-ray luminosity of the two components in the FU\,Tau binary 
compared to other BDs and TTS in Taurus. 
Note, that we plot the total X-ray luminosity for FU\,Tau\,A, although below we argue that possibly
only the weaker, hot component is of coronal origin. 
If only the hotter spectral component of FU\,Tau\,A,
which is more similar to the plasma detected from other Taurus BDs, is considered
FU\,Tau\,A is still the X-ray brightest BD in Taurus together with CFHT-BD-Tau\,4.  
The data for the BDs was extracted from \cite{Grosso07.1}, those for the TTS are from the
catalog of \cite{Guedel07.3}. In addition, we add data for new Taurus members identified after 
the publication of the first results from XEST. The X-ray data for these objects are 
taken from \cite{Scelsi07.1} and \cite{Mokler02.1}, and their stellar parameters are 
listed by \cite{Luhman09.2}. 
Recall, that contrary to some very-low mass (VLM) stars with
high X-ray luminosities, indicated by vertical lines in Fig.~\ref{fig:lx_teff}, 
FU\,Tau\,A did not show variability. 

\begin{figure}
\begin{center}
\includegraphics[width=9cm]{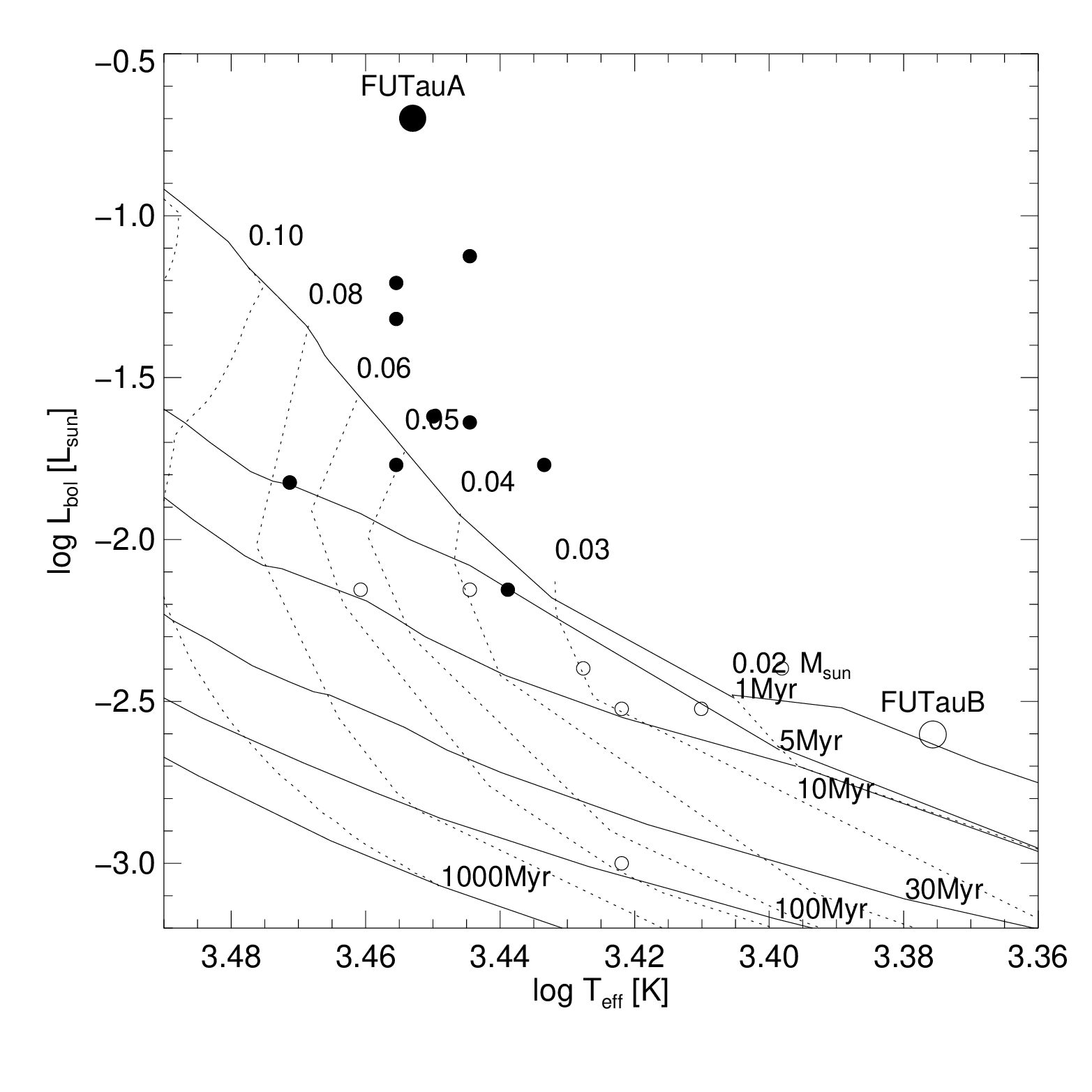}
\caption{HR diagram for VLM objects in Taurus including BDs from \protect\cite{Grosso07.1} and 
the FU\,Tau binary on pre-main sequence models from \protect\cite{Baraffe98.1} and \protect\cite{Chabrier00.2}. 
X-ray detections are shown as filled circles, non-detections are open circles.}
\label{fig:hrd}
\end{center}
\end{figure}

The fractional X-ray luminosity of FU\,Tau\,A ($\log{L_{\rm x}/L_{\rm bol}} = -3.3$ in $0.5-8$\,keV) 
is within the range observed for other Taurus BDs. 
If only the hot spectral component of FU\,Tau\,A is considered
$\log{(L_{\rm x}/L_{\rm bol})} = -3.7$ 
which is even more typical for young BDs; the median for the XEST BDs is $-4.0$
according to \cite{Grosso07.1}. 
This `normal' $L_{\rm x}/L_{\rm bol}$ ratio despite the high $L_{\rm x}$ is a consequence of the 
particularly high bolometric luminosity of FU\,Tau\,A. 
\cite{Luhman09.1} give $L_{\rm bol} = 0.2 L_\odot$, 
about a factor $3$ higher than for the optically brightest of the BDs examined by \cite{Grosso07.1}. 
Its position more than $1$\,dex 
above the youngest isochrone of the evolutionary calculations by \cite{Baraffe98.1}
can be seen in Fig.~\ref{fig:hrd}. While several other BDs in Taurus are located above the $1$\,Myr isochrone
as well, FU\,Tau\,A is clearly an extreme case. 

\subsection{X-ray emission from substellar twins}\label{subsect:disc_twins}

The components in a binary are usually believed to be coeval. For BD binaries resolvable with {\em Chandra} 
this enables a study of X-ray properties from two substellar objects of different spectral type, i.e. mass.
In the common hypothesis that the X-ray emission from BDs is produced in a corona
this means investigating the temperature and mass dependence of magnetic activity. 

We detected the optically brighter primary, FU\,Tau\,A, with {\em Chandra} 
at 
more than two decades higher 
count rate 
than the upper limit for the undetected secondary. 
Considering that these two BDs differ by only two spectral 
subclasses, our result thus implies either (i) a very steep decrease of X-ray 
activity between spectral type M7 and M9 or 
(ii) a crucial influence of some other parameter on the efficiency of magnetic activity or 
(iii) a misinterpretation of the link between the two stars, or 
(iv) an altogether different X-ray emission process. 
Next these hypotheses are investigated. 

Fig.~\ref{fig:lx_teff}, indeed, points at a steep drop in X-ray efficiency around
$T_{\rm eff} \sim 2700$\,K. However, the decay of $L_{\rm x}$ goes along with decreasing optical
brightness of BDs, and the sensitivity of XEST and our {\em Chandra} data for FU\,Tau do not allow to
ascertain whether systematic changes of $L_{\rm x}/L_{\rm bol}$ occur in the substellar regime at the
age of Taurus (few Myr). 

Extinction is unlikely the reason for the non-detection
of FU\,Tau\,B because its $A_{\rm V}$ is smaller than that of component A; see however \cite{Stelzer07.1} for 
the case of the BD binary 2MASS\,J11011926-7732383\,AB where the component with higher $A_{\rm V}$ was brighter
in X-rays than the one with lower $A_{\rm V}$. 
Possible parameters that rule magnetic activity are the rotation rate and the magnetic field strength and 
structure. 
It was shown that pre-MS stars are in the
saturated regime of the activity-rotation relation 
where $L_{\rm x}$ is independent of rotation \citep{Preibisch05.1, Briggs07.1}.
While {\em evolved} VLM stars show little or no X-ray emission despite rapid rotation \citep{Berger10.1}, 
the connection between X-rays and rotation has not yet been examined for {\em young} objects in  
the substellar regime. 
Rotation periods from few hours to several days have been measured for young 
BDs \citep[e.g.][]{Scholz04.3, Joergens03.1}.
Rotation and magnetic field are unexplored for FU\,Tau; see Sect.~\ref{subsect:disc_hrd} for an estimate of both. 

Another possibility that would explain the strongly different X-ray emission levels of FU\,Tau\,A and~B is 
that the two objects are {\em not} coeval with the primary being younger and, therefore, more active. 
This interpretation is in line with the HR diagram position of FU\,Tau\,A {\em above} the models while FU\,Tau\,B 
coincides with the $1$\,Myr isochrone.
An initial age difference of $\leq 1$\,Myr would become impossible to recognize within a few Myrs. 
Alternatively, FU\,Tau\,A and~B may be unrelated objects, with component A being much closer than the assumed
$140$\,pc. This would explain its high X-ray and bolometric luminosity. However,  
the binary nature of this BD system was established by \cite{Luhman09.1} on the basis of the low
probability for a chance projection, especially considering the high extinction through the B\,215 cloud. 
 
Finally, the different X-ray properties of FU\,Tau\,A and~B might point at different emission mechanisms 
in the two objects. 
In Sect.~\ref{subsect:disc_twhya} we suggest 
that a major part of the X-ray emission from FU\,Tau\,A may be related
to the accretion process and not a confined corona.

\subsection{FU\,Tau\,A, a substellar analog to TW\,Hya?}\label{subsect:disc_twhya}

\begin{figure}
\begin{center}
\resizebox{8cm}{!}{\includegraphics{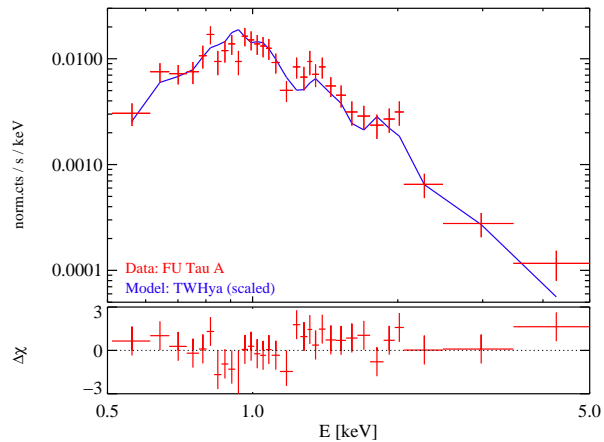}}
\caption{
{\em Chandra}/ACIS spectrum of FU\,Tau\,A 
and residuals with respect to the bestfit presented in Table~\ref{tab:xspec} (red in color print) 
compared to the 2-T bestfit model for TW\,Hya (solid line; blue in color print). 
The spectrum of TW\,Hya has been scaled by a factor of $0.07$ to make the comparison with
FU\,Tau\,A easier. We also have artificially increased the absorbing column density of TW\,Hya 
to the value measured for FU\,Tau\,A and reduced the neon abundance from its atypically high
value to solar. 
}
\label{fig:acis_spectrum}
\end{center}
\end{figure}

In Fig.~\ref{fig:acis_spectrum} we display the X-ray spectrum of FU\,Tau\,A. 
As mentioned already in Sect.~2.1, the spectral fitting revealed an unexpectedly strong soft plasma component.
This characteristic is reminiscent of the TTS TW\,Hya. 
In Table~\ref{tab:xspec} we compare the best-fitting parameters 
of FU\,Tau\,A to those obtained from an {\em XMM-Newton}/EPIC observation for TW\,Hya.  
Despite the best fit result 
presented by \cite{Stelzer04.3} for TW\,Hya was a $3$-T model, we show here the parameters for 
a $2$-T model for better comparison with FU\,Tau\,A. 
This model spectrum for TW\,Hya is overlaid in Fig.~\ref{fig:acis_spectrum} on the observed spectrum of 
FU\,Tau\,A. (A direct graphical comparison of the two observations is impossible because of the different
instruments with which the data were obtained.) 
To ease the comparison, the model of TW\,Hya was multiplied with a factor $0.07$ 
(accounting for the difference in distance and the factor two difference in X-ray luminosity), 
the column density was increased to the value observed for FU\,Tau\,A, and the neon abundance was decreased
to the solar level. 
There is almost a $1:1$ correspondence between the spectrum of FU\,Tau\,A and the down-scaled, 
artificially absorbed spectrum of TW\,Hya, demonstrating that the temperature structure is very similar. 
%

Both objects are dominated by soft emission with only a weak emission component from 
the higher energy photons that are typical for stellar coronae. 
By means of high-resolution X-ray spectroscopy, plasma densities ($n_{\rm e} \sim 10^{12}\,{\rm cm^{-3}}$) 
two dex higher than in stellar coronae have been measured for 
TW\,Hya \citep{Kastner02.1, Stelzer04.3}. 
Its soft X-ray emission 
and the measured high densities are compatible with the conditions in an accretion shock.  
This way TW\,Hya was established as the prototype of a pre-MS star with X-ray emission from accretion columns.   
A handful of other accreting TTSs have since been recognized
to display high density X-ray emitting plasmas \citep[e.g.][]{Schmitt05.1, Guenther06.1}, and an excess
of soft X-rays over the typical coronal emission of young stars has been detected in a number of accreting TTSs 
\citep{Guedel07.4}.
Obtaining a high-resolution X-ray spectrum for FU\,Tau\,A or any other BD 
is out-of-reach for present-day X-ray instrumentation, such that its plasma density can not be constrained. 
However, on the basis of the similar temperature structure between the X-ray spectrum of FU\,Tau\,A and TW\,Hya
we speculate that the soft X-ray emission from FU\,Tau\,A, represented by 
the dominating component\,\#\,1 in the spectral fit, may be produced in accretion shocks rather than in a corona.

To check if the accretion scenario is compatible with the observed properties we compute the 
theoretically expected emission measure from the post-shock region, 
\begin{equation}
EM_{\rm psh} = \dot{M}_{\rm acc} \cdot \frac{3 k_{\rm B} \cdot T_{\rm psh}}{\mu m_{\rm p} P(T_{\rm psh})}
\label{eq:em_theo}
\end{equation}
where $k_{\rm B}$ is the Boltzmann constant, $\mu = 0.61$ the mean particle weight in units of 
the proton mass $m_{\rm p}$, 
and $P(T_{\rm psh})$ the radiative loss function extracted from APEC \citep{Smith01.1}. 
The post-shock temperature is $T_{\rm psh} = \frac{3}{16} \frac{\mu m_{\rm p}}{k_{\rm B}} v^2_{\rm 0}$, 
where $v_{\rm 0}$ is the infall speed. 
For the accretion rate, $\dot{M}_{\rm acc}$, we substitute the value obtained from He\,$5876$, 
$\dot{M}_{\rm He\,I}$. 
Adopting for $v_{\rm 0}$ the velocities measured from the width of the H$\alpha$ profile, 
$v_{\rm H\alpha} \sim 175$\,km/s, 
we find $kT_{\rm psh} \sim 0.03$\,keV and 
$\log{EM_{\rm psh}}\,{\rm [cm^{-3}]}  = 52.7$. While the predicted emission measure is roughly 
comparable to the soft X-ray plasma, $EM_1$ from Table~\ref{tab:xspec}, the post-shock temperature is 
smaller than the observed X-ray temperature, $kT_1$. 
The thermal emission connected to the low temperature expected from material infalling towards a BD should, 
indeed, be too soft to be significant in the X-ray band. On the other hand, as discussed in 
Sect.~\ref{subsect:disc_bds} the dominance of the soft
X-ray component is completely untypical for magnetic activity on such a {\em young} BD. 
So, if accretion is not responsible for FU\,Tau\,A's X-ray emission, it must have a very unusual
corona. 

\subsection{Accretion properties of FU\,Tau\,A}\label{subsect:disc_mdot}

FU\,Tau\,A stands out among BDs of similar mass for its high accretion rate. 
The empirical $M - \dot{M}_{\rm acc}$ relation has a large spread of controversial origin. From optical spectroscopy, 
values of $\dot{M}_{\rm acc} \sim 10^{-12...-9}\,{\rm M_\odot/yr}$ 
have been measured for BDs with $M \sim 0.05\,M_\odot$ \citep[e.g.][]{Natta06.1}.  
However, in that mass range accretors with $\dot{M}_{\rm acc} \geq 10^{-10}\,{\rm M_\odot/yr}$ are found only
in $\rho$\,Oph, the youngest of the examined star-forming sites \citep{Natta04.2}, and 
the $M_* - \dot{M}_{\rm acc}$ relation predicts $\log{\dot{M}_{\rm acc}}\,{\rm [M_\odot/yr]} = -10.7$,
more than $1$\,dex below our value for $\dot{M}_{\rm He}$. 

Strong accretion produces hot spots on the surface of the star that might introduce
brightness variations in the course of a rotation cycle. 
The two epochs of {\em Sloan Digital Sky Survey} (SDSS) photometry (from Dec 6 and 29 2002) presented
by \cite{Luhman09.1}, indeed, indicate strong flux variations. 
The amplitude is $0.07$\,mag in the $z$ band and  
increases monotonously towards blue wavelengths (e.g., $0.27$\,mag in  
$r$ band, $0.44$\,mag in $g$ band). 
The wavelength dependence from a standard extinction law \citep{Mathis90.1} is not 
steep enough to explain this behaviour with variable extinction and we assume
that we see the signature of star spots. 
Using a simple spot model comparable to the one presented by \cite{Scholz05.2} 
and assuming blackbody spectra for the spotted and unspotted surface area, we find that 
cool spots with $\Delta T = 0-1000$\,K and filling factor $f = 5-50$\,\% do not provide 
the observed steep decrease of amplitude with wavelength. 
On the other hand, hot spots (caused by the accretion flow) with $\Delta T > 500$\,K  
and $f = 5-10$\,\% yield a decent match to the observed spectral dependence  
of the amplitudes. 
We assume here that the minimum photometry corresponds to the photosphere which may not be the case. 
Clearly, the two epochs of SDSS data
provide only a lower limit to the variability of FU\,Tau\,A. Spots with $\Delta T$ of a few $1000$\,K,
as have been measured on other BDs \citep{Scholz09.1}, 
can not be ruled out for FU\,Tau\,A on the basis of existing photometry. 
Hot spots with these characteristics have a 
luminosity of 
$L_{\rm spot} = 1.5 \cdot 10^{32}$\,erg/s, i.e $\sim 20$\,\% of the bolometric luminosity of FU\,Tau\,A. 

\subsection{The relation between stellar parameters and accretion/activity in FU\,Tau\,A}\label{subsect:disc_hrd}

One puzzling feature of FU\,Tau\,A is its position high above the youngest isochrone in the HR diagram. 
An accurate assessment of (sub)stellar parameters is crucial for
the evaluation of accretion and X-ray properties as can be seen from the following open problems.  
First, the velocities derived in Sect.~\ref{subsect:disc_twhya}
from the H$\alpha$ line 
are difficult to reconcile with the free-fall speed 
for the published (sub)stellar parameters of FU\,Tau\,A. In particular its large $L_{\rm bol}$
and ensuing huge radius ($1.8\,R_\odot$) imply $v_{\rm ff} \sim 100$\,km/s, almost a factor
two lower than the value derived from H$\alpha$.
Secondly,  
the spot luminosity and the luminosity derived from the kinetic energy of the infalling material 
($L_{\rm acc, He\,I}$) could be better reconciled 
assuming a $1$ dex fainter $L_{\rm bol}$ for FU\,Tau\,A;  
this would moreover make the BD compatible with the $1$\,Myr isochrone.
On the other hand, the high bolometric luminosity of FU\,Tau\,A comes 
along with high X-ray luminosity, 
supporting an unexplained trend first 
noticed by \cite{Stelzer07.3} who found that most of the BDs in star-forming regions 
detected in X-rays are located on the $1$\,Myr isochrone or above the evolutionary calculations. 
This trend is also seen in Fig.~\ref{fig:hrd}. 
Thus, FU\,Tau\,A may not be an isolated albeit an extreme case.  
In the following we consider various possible causes for the high luminosity of FU\,Tau\,A: 
magnetic activity, accretion, problems with models and wrong distance.

\subsubsection{Reduced convection affecting temperature and radius}\label{subsubsect:disc_hrd_activity}

A mechanism that can produce larger radius (and cooler temperature) for a given mass,
without affecting the bolometric luminosity, i.e. shifting the object horizontally 
in the HR diagram,  
is reduced convection efficiency and ensuing heat flux because of fast rotation and/or strong 
magnetic fields.
Such a scenario has been investigated by \cite{Chabrier07.1} and by \cite{MacDonald09.1}
using different approaches to model the influence of the magnetic field.
Both studies have successfully explained the temperature reversal of the eclipsing spectroscopic
binary 2M\,05352184-0546085 \citep{Stassun07.2} with strong magnetic effects on the primary. 
Indeed, in that object the primary which has a lower effective
temperature than the secondary is strongly active in terms of high rotation velocity and strong H$\alpha$
emission \citep{Reiners07.3}. 

Shifting FU\,Tau\,A to higher temperatures onto the $1$\,Myr isochrone in the HR diagram
to make up for a presumed effect of activity on $T_{\rm eff}$ and $R_*$ 
yields a `true' mass of $\sim 0.2\,M_\odot$ and `nominal' temperature of $\sim 3190$\,K. 
We can estimate the rotation period and field strength required for 
magnetospheric accretion models. We adopt the X-wind model that defines the disk truncation radius
$R_{\rm x}$ as the distance from the star where the Keplerian angular velocity 
equals the stellar angular velocity  
\citep[see][]{Ostriker95.1, Mohanty08.1}. Assuming $R_{\rm in} = 2\,R_{\rm *}$ 
we obtain with the radius of $1.83\,R_\odot$ derived from luminosity and temperature and with the 
`corrected' stellar mass of FU\,Tau\,A a period of 
$P_{\rm rot} \sim 1.8$\,d. Using the model of \cite{Koenigl91.1} where the disk
truncation radius is smaller than the corrotation radius yields a longer period, 
$P_{\rm rot} \sim 5$\,d. 
We note in passing that a larger inner disk radius 
($R_{\rm in} = 5\,R_{\rm *}$ is often used in the literature) 
yields 
periods between $7...21$\,d depending on the details of the accretion model. 
These latter values for the period are at the high end of the values 
measured for BDs and TTS \citep[e.g.][]{Scholz04.3, Herbst07.2}. Indeed, 
\cite{Rebull06.1} presented evidence that stars with disks have longer periods on a statistical
basis than stars without disks but this does not exclude the existence of (some) fast-rotating accretors. 
For the observed $\dot{M}_{\rm He\,I}$, 
\cite{Koenigl91.1} predicts for FU\,Tau\,A a 
magnetic field between $B \sim 85...400$\,G, for assumptions of 
$R_{\rm in} = 2\,R_{\rm *}$ and $5\,R_\odot$, respectively.  
Magnetic fields with kG strength are not unusual for TTS \citep{JohnsKrull07.2}.
For BDs with ages of $\sim 1-10$\,Myr \cite{Reiners09.2} could not find detectable fields 
with upper limits of $\leq 1$\,kG for all of the accretors, 
while young but non-accreting BDs and evolved ultracool dwarfs in the field exhibit field strengths 
of a few kG \citep{Reiners07.2}. 
However, the sample of only three accretors studied by \cite{Reiners09.2} is too small
to be clear evidence against the presence of $\leq 1$\,kG fields on accreting substellar objects. 

To summarize, for a reasonable range of $R_{\rm in}$ we obtain either fast 
rotation or strong magnetic field for FU\,Tau\,A, 
i.e. at least one of the two conditions for suppressing convection seems 
to be fullfilled independent of the position of the disk truncation radius. 
The high level of (X-ray) activity of FU\,Tau\,A -- 
even neglecting the dominating cool X-ray plasma possibly produced by accretion -- 
is compatible with this scenario. 
Similarly, the high value for the H$\alpha$ flux corresponds to a high $L_{\rm H\alpha}/L_{\rm bol}$
ratio. However, given the dominant contribution of accretion in forming the H$\alpha$ line this
measurement can not be used as a diagnostic for strong magnetic activity in FU\,Tau\,A, contrary to the
case of the non-accreting eclipsing binary 2MASS\,05352184-0546085 \citep{Reiners07.3}.
Cool star spots are expected as a manifestation of inhibited convection. 
As argued in Sect.~\ref{subsect:disc_mdot}, the two epochs of photometric observations can not
be explained by cool spots alone, but they can not rule out the presence of cool spots either.
More detailed monitoring is required to test this scenario. 

While the above considerations are not conclusive, in the absence of measured rotation, field
strength and good constraints on the nature of the spots it is 
not excluded that rotation/fields affect the stellar parameters of FU\,Tau\,A. 
However, the presence of accretion, contrary to the case of 2M\,05352184-0546085, certainly complicates the 
situation. 
A quantitative theoretical evaluation of magnetic field effects in this particular object
would certainly provide further insight. 
We now discuss a few implications of this scenario for the interpretation of our X-ray and H$\alpha$
observations. First, the higher mass would move
FU\,Tau\,A into the stellar regime. Higher mass objects have higher X-ray luminosity, i.e. 
FU\,Tau\,A would be less extreme in terms of X-ray brightness. 
The $L_{\rm x}/L_{\rm bol}$ ratio would
be unchanged and it is, indeed, compatible with the values observed in young VLM stars; in the Orion
Nebula Cluster the median of $0.2\,M_\odot$ stars is $\log{(L_{\rm x}/L_{\rm bol})} \sim -3.5$ and the 
spread is at 
least one logarithmic dex \citep{Preibisch05.1}. 
Secondly, the free-fall velocity would be higher ($\sim 200$\,km/s) 
and in better agreement with the observed H$\alpha$ profile. Moreover, the higher velocity  
corresponds to a predicted post-shock temperature which is more similar to the X-ray observed temperature. 
Finally, the accretion luminosity derived from $\dot{M}_{\rm He}$ would increase by a factor four
but still be much smaller than the estimated spot luminosity.

\subsubsection{Excess accretion luminosity}

Alternatively, the position of in the HR diagram for FU\,Tau\,A and similar objects may be 
due to excess luminosity related to accretion. \cite{Luhman09.1} have
estimated $L_{\rm bol}$ for FU\,Tau\,A from its $J$ band magnitude, where deviations from photospheric emission 
due to the presence of a disk and accretion are usually smallest.  
However, strong (accretion) variability in the $J$ band can not be ruled out with the available data. 

If the true bolometric luminosity of FU\,Tau\,A is significantly lower than the adopted value
the radius is smaller (for given mass of $0.05\,M_\odot$) 
and the free-fall velocity gets larger, 
e.g. a factor $10$ smaller $L_{\rm bol}$ gives $v_{\rm ff} = 180$\,km/s
in perfect agreement with the observed H$\alpha$ width. 
On the other hand, this interpretation 
implies a high fractional X-ray luminosity for FU\,Tau\,A and the other BDs that are located 
above the models in the HR diagram ($L_{\rm x}/L_{\rm bol} \geq 10^{-3}$).
Such high X-ray emission levels are difficult to explain in terms of T\,Tauri like activity. 
Moreover, two of the three Taurus BDs that are closest in the HR diagram to FU\,Tau\,A,
i.e. highest above the models, are not accreting based on the equivalent width of their H$\alpha$
emission \citep{Grosso07.1} and they even have no disks according to \cite{Luhman10.1}, making 
accretion luminosity an unlikely explanation for their discrepancy with the evolutionary models.

\subsubsection{Shortcomings of evolutionary models}

It is well-known that different sets of pre-main sequence models are inconsistent among each other and
none of them describes adequately all observations. 
It is not clear to what extent these discrepancies between data and model are due to problems with 
the stellar parameters derived from observations or due to shortcomings in the theory. 
\cite{Baraffe09.1} showed that the typically observed age spread of $1-10$\,Myr 
in the HR diagram for stellar populations in star
forming regions can be explained by episodic accretion during the protostellar
phase. However, they 
can not explain the presence of objects above the $1$\,Myr isochrone. 
These authors 
suggest that such objects are very young, having experienced their episodic protostellar accretion
events quite recently and not yet contracted to the $1$\,Myr position of non-accreting objects of 
the same mass. In this view the two components of the FU\,Tau binary are not coeval.  

An age difference of a few hundred thousand years
has been detected for the first time in an equal-mass eclipsing binary, Par\,1802, by \cite{Stassun08.1},
demonstrating that subsequent formation of the components in stellar systems is possible. 
For the case of FU\,Tau the existing evolutionary models are not helpful for constraining
the possible age difference.  
Compared to Par\,1802, FU\,Tau is of lower mass and much wider separation. Moreover, the masses of the 
two components are not equal introducing additional uncertainties in an evalution of their evolution.   
To our knowledge, star formation theories have not yet made predictions on such young non-coeval binaries, 
their properties and possible influence of the environmental conditions.

\subsubsection{Wrong distance or binarity of FU\,Tau\,A}

Finally, we mention for completeness further scenarios 
that would reduce the bolometric luminosity of FU\,Tau\,A 
and bring it in accordance with the $1$\,Myr isochrone: 
(i) A much closer distance than assumed; 
this would also result in a lower, and more typical, X-ray luminosity. 
However, the ensuing non-binarity with component\,B is countered by statistical arguments,
and distance is unlikely to be responsible for similar cases of other BDs above the evolutionary models. 
(ii) The primary in the FU\,Tau system might be a binary; however, 
this would not provide a solution because it would decrease $L_{\rm bol}$ by at most a factor of two.

\section{Summary}\label{sect:conclusions}

Among the most puzzling features in the enigmatic BD FU\,Tau\,A are 
a very high bolometric luminosity for its effective temperature and its strong and unexpectedly soft
X-ray luminosity pointing at an origin in accretion rather than a corona. 

Assuming a mis-interpretation of the stellar parameters as discussed in Sect.~\ref{subsect:disc_hrd} 
would resolve many open problems: HR diagram position, free-fall versus observed velocities, 
spectroscopic and photometric accretion luminosities. 
Suppressed convection as a result of fast rotation and/or strong magnetic field might
be responsible for the apparent excess luminosity of FU\,Tau\,A with respect to pre-main sequence
evolution calculations. If confirmed this would be the first case of an {\em accreting} BD
with strong field effects. 
However, in the absence of observational constraints on the rotation rate
and field strength this explanation is not much more than a plausible hypothesis. 
Keeping in mind the many pieces of evidence for strong accretion in FU\,Tau\,A from 
optical photometry, optical spectroscopy and X-ray emission, 
it is possible that the excess luminosity may be explained by 
a significant contribution of accretion to the $J$ magnitude.
Further constraints on the accretion properties 
of this benchmark BD are required to give our estimates a quantitatively sound basis. 
Last but not least, the existence of other BDs that have stellar parameters incompatible with evolutionary
models, albeit less pronounced, may suggest problems with the ages predicted by evolutionary models.
If FU\,Tau\,A is younger than $1$\,Myr and not coeval with FU\,Tau\,B, as suggested by \cite{Baraffe09.1},
its extraordinarily strong and soft X-ray emission, irrespective of whether due to accretion or to activity,
could then be a result of its young age. 
In the near future we aim at further constraining observationally the 
accretion and activity of this unusual object. 

Our {\em Chandra} observations of FU\,Tau\,A 
represent the first indications for X-ray emission from accretion shocks in a BD
(provided FU\,Tau\,A {\em is} a substellar object and not a star as argued in 
Sect.~\ref{subsubsect:disc_hrd_activity}). 
We note, that even if free-fall velocity and H$\alpha$ width of FU\,Tau\,A
can be reconciled with the scenarios described in Sect.~\ref{subsect:disc_hrd}, the predicted X-ray
post-shock temperature is still lower than our observed value. While this discrepancy remains
unexplained, a similar case is represented by the TTS Hen\,3-600. This object shows clear evidence
of accretion from its H$\alpha$ characteristics as well as the density and softness of its X-ray emitting
plasma \citep{Huenemoerder07.1}. However, quantitatively its mass ($0.2\,M_\odot$) and radius 
($R = 0.9\,R_\odot$) imply a post-shock temperature of $1.1$\,MK, to be compared to the measured X-ray
plasma temperature of $\sim 3$\,MK.

\section*{Acknowledgments}

We would like to thank the referee, S.Mohanty, for his careful reading and constructive comments
that have stimulated us to present a more coherent and comprehensive 
discussion of all plausible interpretations.
This research has made use of data obtained from {\em Chandra} (Obs-ID\,10984) and software provided by the 
{\em Chandra} X-ray Center (CXC) in the application package CIAO. 

\bibliographystyle{mn2e} 
\bibliography{mnemonic,futau}

\label{lastpage}

\end{document}